\documentclass[12pt]{article}
\usepackage{epsfig}
\usepackage{times}
\date{}

\begin{document} 

\title{{\LARGE\sf Local vs.~Global Variables for Spin Glasses}}
\author{
{\bf C. M. Newman}\thanks{Partially supported by the 
National Science Foundation under grant DMS-01-02587.}\\
{\small \tt newman\,@\,cims.nyu.edu}\\
{\small \sl Courant Institute of Mathematical Sciences}\\
{\small \sl New York University}\\
{\small \sl New York, NY 10012, USA}
\and
{\bf D. L. Stein}\thanks{Partially supported by the 
National Science Foundation under grant DMS-01-02541.}\\
{\small \tt dls\,@\,physics.arizona.edu}\\
{\small \sl Depts.\ of Physics and Mathematics}\\
{\small \sl University of Arizona}\\
{\small \sl Tucson, AZ 85721, USA}
}

\maketitle

\begin{abstract}
We discuss a framework for understanding why spin glasses differ so
remarkably from homogeneous systems like ferromagnets, in the context of
the sharply divergent low temperature behavior of short- and infinite-range
versions of the same model.  Our analysis is grounded in understanding the
distinction between two broad classes of thermodynamic variables -- those
that describe the {\it global\/} features of a macroscopic system, and
those that describe, or are sensitive to, its {\it local\/} features.  In
homogeneous systems both variables generally behave similarly, but this is
not at all so in spin glasses.  In much of the literature these two
different classes of variables were commingled and confused.  By analyzing
their quite different behaviors in finite- and infinite-range spin glass
models, we see the fundamental reason why the two systems possess very
different types of low-temperature phases.  In so doing, we also reconcile
apparent discrepancies between the infinite-volume limit and the behavior
of large, finite volumes, and provide tools for understanding inhomogeneous
systems in a wide array of contexts. We further
propose a set of `global variables' that are definable and sensible for
both short-range and infinite-range spin glasses, and allow a meaningful
basis for comparison of their low-temperature properties.

\end{abstract}

{\bf KEY WORDS:\/} spin glass; Edwards-Anderson model;
Sherrington-Kirkpatrick model; replica symmetry breaking; mean-field
theory; pure states; metastates; domain walls; interfaces

\vfill\eject

\small
\renewcommand{\baselinestretch}{1.25}
\normalsize

\section{Local variables and thermodynamic limits}
\label{sec:local}

In recent years, attempts have been made~\cite{MPR97,MPRRZ00} to draw a
distinction between the thermodynamic limit as a `mathematical tool' of
limited physical relevance, and the physical behavior of large, finite
systems, the real-world objects of study.  In this note we discuss why this
distinction is spurious, and show through several examples that this `tool'
is useful precisely {\it for\/} determining the behavior of large, finite
systems.  We will also discuss why this attempted distinction has caused
confusion in the case of spin glasses, and how resolving it introduces some
important new physics.  For ease of discussion, we confine ourselves
throughout to Ising spin systems.

One of the incongruities of the `thermodynamic limit vs.~finite volume'
debate is that true thermodynamic states are in fact measures of the {\it
local\/} properties of a macroscopic system, while discussions of
finite-volume properties --- at least in the context of infinite-range spin
glasses --- focus entirely on {\it global\/} quantities.

Of course, traditional thermodynamics (as opposed to statistical mechanics)
is entirely a study of global properties of macroscopic systems.
Quantities like energy or magnetization are collective properties of {\it
all\/} of the individual spins in a given 
finite- or infinite-volume configuration.  Other global measures cannot be
discussed in terms of individual spin configurations but rather are
meaningful only in the context of a Gibbs distribution (also known as Gibbs
state or thermodynamic state --- we will use these terms interchangably
throughout).  Entropy is the obvious example of such a variable.  All of
these together --- energy, entropy, magnetization, and the various free
energies associated with them --- convey only coarse information about a
system (though still extremely valuable).

What they convey little information about is the actual spatial structure
of a state, or of the relationships among different states.  Even
a quantity like the staggered magnetization, which gives some information
about spatial structure, does not shed significant light on local
properties.

But for real systems one often does want information about local
properties.  In order to analyze local spatial and temporal structures one
generally needs to employ {\it local thermodynamic variables\/} --- for
example, $1-,2-,\ldots,n-$point correlation functions.  In fact, these
functions taken altogether convey {\it all\/} of the information that can
be known about that state in equilibrium --- a Gibbs state is a
specification of all possible $n$-point (spatial) correlation functions,
for every positive integer $n$.

Alternatively, one can define a thermodynamic state either as a convergent
sequence (or subsequence) of finite-volume states as volume tends to
infinity, or else intrinsically through the DLR equations~\cite{Georgii88}.
But we will avoid a technical discussion here in the interest of keeping
the discussion focused on physical objects.  We henceforth assume
familiarity with concepts such as thermodynamic mixed state and
thermodynamic pure state, which have been used extensively throughout much
of the spin glass literature, and refer the reader who wishes to learn more
to Section 4 of~\cite{NS03jpc}.

Are there any {\it global\/} quantities that say something about the
spatial structure of a state?  The answer is yes, and one in particular has
proven to be very useful in comparing different pure state structures in
the infinite-range Sherrington-Kirkpatrick~(SK) model~\cite{SK75} of
a spin glass.  The SK Hamiltonian (in volume $N$) is
\begin{equation}
\label{eq:SK}
{\cal H}_N=-(1/\sqrt{N})\sum_{1\le i<j\le N} J_{ij} \sigma_i\sigma_j 
\end{equation}
where the couplings $J_{ij}$ are independent, identically distributed
random variables chosen, e.g.,  from a Gaussian distribution with zero mean and
variance one; the $1/\sqrt{N}$  scaling ensures a sensible thermodynamic limit
for free energy per spin and other thermodynamic quantities.

In a series of papers, Parisi and
collaborators~\cite{P79,P83,MPSTV84a,MPSTV84b} proposed, and worked out the
consequences of, an extraordinary {\it ansatz\/} for the nature of the
low-temperature phase of the SK model.  Following the mathematical
procedures underlying the solution, it came to be known as {\it replica
symmetry breaking\/} (RSB).  The starting point of the Parisi solution was
that the low-temperature spin glass phase comprised not just a single
spin-reversed pair of states, but rather ``infinitely many pure
thermodynamic states''~\cite{P83}, not related by any simple symmetry
transformations.

But how were they related?  To answer this question, Parisi introduced a
{\it global\/} quantity --- exactly of the sort we were just asking about
--- to quantify such relationships.  The actual notion of pure `state' in
the SK model is problematic, as discussed, e.g.,
in~\cite{NS03jpc,NSBerlin,NS03mf,Ta03}.  We'll ignore that problem for now,
though, and assume that somehow two SK pure states $\alpha$ and $\beta$
have been defined.  Then their overlap $q_{\alpha\beta}$ is defined as
\begin{equation}
\label{eq:qabSK}
q_{\alpha\beta}= {1\over N}\sum_{i=1}^N\langle\sigma_i\rangle_\alpha\langle\sigma_i\rangle_\beta\, ,
\end{equation}
where $\langle\cdot\rangle_\alpha$ is a thermal average in pure state
$\alpha$, and dependence on ${\cal J}$ and $T$ has been suppressed.  So
$q_{\alpha\beta}$ is a quantity measuring the similarity between states $\alpha$
and $\beta$.  

We noted above that quantities referring to individual pure states are
problematic in the SK model, since there is no known procedure for
constructing such states in a well-defined way.  However, what is really of
interest is the {\it distribution\/} of overlaps, which {\it can\/} be
sensibly defined by using the finite-$N$ Gibbs state.  The overlap
distribution is constructed by choosing, at fixed $N$ and $T$, two of the
many pure states present in the Gibbs state.  The probability that their
overlap lies between $q$ and $q+dq$ is then given by the quantity $P_{\cal
J}(q)dq$, where
\begin{equation}
\label{eq:ovdistSK}
P_{\cal J}(q)=\sum_\alpha\sum_\beta W_{\cal J}^{\alpha}W_{\cal J}^{\beta}\delta(q-q_{\alpha\beta})\, .
\end{equation}
As before, we suppress the dependence on $T$ and $N$ for ease of notation.
The average $P(q)$ of $P_{\cal J}(q)$ over the disorder distribution is
commonly referred to as the {\it Parisi overlap distribution\/}, and serves
as an order parameter for the SK model.

Because there is no spatial structure in the infinite-range model, the
overlap function does seem to capture the essential relations among the
different states.  However, it might already be noticed that such a global
quantity would miss important information in short-range models ---
assuming that such models also have many pure states.  There is no
information in $P_{\cal J}(q)$ about local correlations.  This is
acceptable, even desirable, in an infinite-range model such as SK which has
no geometric structure, measure of distance, or notion of locality or
neighbor.  But all of these are well-defined objects in short-range models,
and carry a great deal of information about any state, pair of states, or
collection of many states.  This is one of the sources of the difficulties
one encounters (see, e.g.,~\cite{NS03mf}) when attempting to apply
conclusions to short-range spin glasses that were derived for the SK model.

\section{Nearest-Neighbor Ising Ferromagnets}
\label{sec:nnifs}

To illustrate some of these ideas in a simple context, consider the uniform
nearest-neighbor Ising ferromagnet on ${\bf Z}^d$, with Hamiltonian
\begin{equation}
\label{eq:ferroham}
{\cal H}=-\sum_{{x,y}\atop|x-y|=1}\sigma_x\sigma_y\, .
\end{equation} 

It is natural in models such as this to take periodic or free boundary
conditions when considering the finite-volume Gibbs state.  In any fixed
$d\ge 2$, consider for $T<T_c$ a sequence of volumes $\Lambda_L$ (for
specificity, $L^d$ cubes centered at the origin) tending to infinity with,
for example, periodic b.c.'s.  The Gibbs state --- which, as emphasized in
the preceding section, describes behavior of the local spin variables ---
converges in the infinite-volume limit to the symmetric mixture of a pure
plus state with $\langle\sigma_x\rangle>0$ (which, because of translation
symmetry, is independent of $x$) and a pure minus state with
$\langle\sigma_x\rangle<0$.  In a similar way, one could choose to study
instead a global variable, say the magnetization per spin 
$M_L=|\Lambda_L|^{-1}\sum_{x\in\Lambda_L}\sigma_x$.  It is easy to show
that this variable has a distribution that converges in the limit to a
symmetric mixture of $\delta$-functions at $\pm\langle\sigma_x\rangle$.  In
this case descriptions of the system in terms of both local and global
variables are interesting, and more to the point, they agree.

But even in the simple case of the uniform, nearest-neighbor Ising
ferromagnet this need not always be true.  Consider `Dobrushin
boundary conditions'~\cite{Dob}.  These are b.c.'s in which the boundary
spins above the `equator' (a plane or hyperplane parallel to two opposing
faces of $\Lambda_L$ and cutting it essentially in half by passing just
above the origin) are chosen to be plus and the boundary spins at and below
the equator are minus.  Boundary conditions such as these are useful for
studying interface structure in spin models.

Now (below the roughening temperature) translation-invariance is lost (in
one direction), and the local and global variables disagree.  The Gibbs
state is one where $\langle\sigma_z\rangle>0$ for $z>0$ (taking $z=0$ to be
the equator) and $\langle\sigma_z\rangle<0$ when $z\le 0$.  The magnitude
of $\langle\sigma_z\rangle$ will depend on the value of $z$ (though it
remains independent of the coordinates in all transverse directions).
Moreover for $0<T<T_c$ there is additional dimension-dependence of the
behavior of the local variables (related to the roughening transition)
which we will not discuss here.

The magnetization global variable is no longer even interesting.  Its
distribution converges to a $\delta$-function at 0 at all temperatures in
all dimensions.  It therefore conveys very little information about the
nature of the state. One could instead choose a more appropriate global
variable that better matches the boundary conditions,
e.g., an
order parameter such as
\begin{equation}
\label{eq:mtilde}
\tilde M=\lim_{L\to\infty}|\Lambda_L|^{-1}\sum_{x\in\Lambda_L} g(x)\langle\sigma_x\rangle
\end{equation}
where $g(x)=+1$ if $x$ is above the equator and $-1$ if below.

One might also consider a sort of `quasi-global' variable, that looks at
block magnetizations in blocks that are large compared to the unit lattice
spacing but small compared to entire system size $L$.  One could then
examine the `spatial' distribution of the block magnetization as the
location of the block varies through the system. Above $T_c$ this is simply
a $\delta$-function at zero, but below $T_c$ one gets a symmetric mixture
of $\delta$-functions at $\pm\lim_{z\to\infty}\langle\sigma_z\rangle$ for
all $d$. This is still not as sensitive as the actual Gibbs state, 
which can distinguish between the rough and nonrough interfaces (e.g.,
below $T_c$ in $d=2$ compared to $T<T_R$ in $d=3$, where $T_R$ is the 3D
roughening temperature) by having a different expression for the limiting
Gibbs state in the two cases.

\section{Finite- and Infinite-Range Spin Glasses}
\label{sgs}

In the case of spin glass models, we have
found~\cite{NSBerlin,NS03mf,NS96a,NS96b,NS97,NS98,NS02} a much sharper
disparity between finite- and infinite-range models than is the case for
any homogeneous statistical mechanical model of which we are aware.
Consider first the Edwards-Anderson (EA) nearest-nieghbor model~\cite{EA75}
on ${\bf Z}^d$.  Its Hamiltonian in zero external field is given by
\begin{equation}
\label{eq:EA}
{\cal H}=-\sum_{{{x,y}\atop|x-y|=1}} J_{xy}\sigma_x\sigma_y\ ,
\end{equation}
where the nearest-neighbor couplings $J_{xy}$ are defined in exactly the
same way as the $J_{ij}$ in the SK~Hamiltonian~(\ref{eq:SK}).  In this
model thermodynamic pure, mixed, and ground states are standard,
well-defined (see, e.g.,~\cite{NS03jpc,NS02}) objects, constructed according to
well-established prescriptions of statistical
mechanics~\cite{Georgii88,Ruelle,Lanford,Simon,Slawny,Dobrushin,vEvH}.
Local thermodynamic variables therefore convey in principle all of the
essential information about any state.

Global variables such as Parisi overlap functions can be defined for the EA
model as well, but are now very prescription-dependent: for the same
system, very different overlap functions can be obtained through use of
different boundary conditions, or by changing the order of taking the
thermodynamic limit and breaking the replica symmetry.  Because of this,
they may not convey reliable information about the number of states or the
relationships among them.  For a detailed review of these issues and
problems, see~\cite{NS03jpc}.

Turning to the SK model, we find that a unique situation arises.  Pure
states are in principle defined for a fixed realization of {\it all\/} of
the couplings; but in the SK model the {\it physical\/} couplings
$J_{ij}/\sqrt{N}$ scale to zero as $N\to\infty$. As a result, there does
not now exist any known way of constructing thermodynamic pure states in an
SK spin glass.  It has been proposed~\cite{BY86,MPV87} that one way of
defining such objects is through the use of a modified `clustering'
property: if ${\alpha}$ denotes a putative pure state in the SK model, then
one can demand it satisfy:
\begin{equation}
\label{eq:modclustering}
\langle\sigma_i\sigma_j\rangle_\alpha - 
\langle\sigma_i\rangle_{\alpha}\langle\sigma_j\rangle_{\alpha}\ \to 0 
\qquad {\rm as }\quad N\to\infty\quad ,
\end{equation}
for any fixed pair $i,j$, in analogy with the clustering property obeyed by
`ordinary' pure states in conventional statistical mechanics.  At this time
though, there exists no known operational way to construct such an $\alpha$
as appears in~(\ref{eq:modclustering}). But even if such a construction
were available, the definition of pure states
through~(\ref{eq:modclustering}) leads to bizarre conclusions in the SK
model~(see~\cite{NS03mf, NSother}).

Are these problems simply a consequence of infinite-range interactions?
No, because they are absent in the Curie-Weiss model of the uniform
ferromagnet. Though physical couplings scale to zero there also, they
``reinforce'' each other, being nonrandom, so one may still talk about
positive and negative magnetization states --- in analogy with what one
sees in finite-range models --- in the $N\to\infty$ limit (of course, one
can no longer talk of interface states).  So the unique behavior of the SK
model arises (at the least) from the combination of {\it two\/} properties:
coupling magnitudes scaling to zero as $N\to\infty$, and quenched disorder
in their signs.  (Some success has been achieved in defining states in
mean-field Hopfield models --- see, e.g.,~\cite{BG97,BGP97,Kuelske97} and
references therein --- where the correct order parameters are known {\it a
priori\/}.)

Although individual pure states have not so far been (and perhaps cannot
even in principle be) constructed for the SK model, we have nevertheless
proposed methods, based on chaotic size dependence~\cite{NS92}, that can
detect the {\it presence\/} of multiple pure states.  One can then examine
the nature of objects that are analogous to states and that are defined
through local variables, using the usual prescriptions of statistical
mechanics.  However, an analysis~\cite{NS03mf} of the properties of these
state-like objects shows that they behave in completely
unsatisfactory --- in fact, absurd --- ways.  For example, using the
traditional definition of a ground state --- or equivalently, the modified
clustering property of~(\ref{eq:modclustering}) --- one can prove that (for
almost every fixed coupling realization~${\cal J}$) {\it every\/}
infinite-volume spin configuration is a ground state.  That is, as $N$
increases, any fixed finite set of correlation functions cycles through all
of its possible sign configurations infinitely many times.  Of course, this
cannot happen in short-range spin glasses in any dimension, nor in any
other statistical mechanical model based on any sort of physical system.

The upshot is that global variables (like overlap distributions for the
whole system) capture interesting phenomena in the SK model, while local
variables are not so interesting there; in fact, their use can even be
dangerous in drawing conclusions about realistic spin glass models. This
may be because the SK model itself is {\it a priori\/} a global (or at
least nonlocal) model, and does not lend itself even in principle to any
sort of local analysis.

A large body of evidence compiled by the
authors~\cite{NSBerlin,NS03mf,NS96a,NS96b,NS97,NS98,NS02} shows that local
variables and states in the EA model do not behave anything like those in
the SK model.  The same conclusion applies to global variables in the EA
model constructed in close analogy with those from the SK model --- that
is, in a way intended to convey information about states.  This will be
further discussed in~\cite{NSother}.  Consequently, we expect that attempts
to derive conclusions about the EA model in terms of local properties
(i.e., pure state behavior) from the global behavior of the SK model will
not work.

\section{A New Global Order Parameter for Spin Glasses?}
\label{sec:newop}

In this section we consider an interesting speculative question motivated
by a comment of A.~Bovier~\cite{Bov03} that the usual description of Gibbs
measures for short-range models is inadequate for mean field spin glasses.
We have presented a large body of evidence that any quantity, describing
spin glass properties and that is derived from or based on properties of
pure states, cannot connect the behavior of short-range and infinite-range
spin glasses.  But can one construct a new type of global variable that is
meaningful for both short-range and infinite-range spin glass models, and
allows a direct comparison of their properties?  One obvious candidate is a
`global' overlap function not related to pure states; that is, rather than
computing $P_L(q)$ in a `window'~\cite{NS03jpc,NS98,NS02} far from
$\partial\Lambda_L$, one would compute it in the entire volume $\Lambda_L$.
As discussed in~\cite{NS03jpc,NS98,NS02}, the resulting quantity may be
unrelated to pure state structure.  An analogous situation is the
ferromagnet {\it above\/} the roughening temperature (but below $T_c$).
Even though there are no domain wall states, employing Dobrushin boundary
conditions~(see Sec.~\ref{sec:nnifs}) will generate spin configurations
that on very large scales (say, of order $L$) look almost indistinguishable
from those belonging to domain wall states.  But it's unclear whether doing
this generates any useful or nonobvious information.

Similarly (see~\cite{NS98} for a more detailed discussion) there is reason
to doubt whether a global overlap distribution would be any more useful.
In the SK model boundary conditions are not an issue; but in short-range
models, overlaps are potentially very sensitive to them.  One
consequence~\cite{NS03jpc,HF87a} of this sensitivity is that spin overlap
functions are unreliable indicators of pure state multiplicity: they can
possess a trivial structure (cf.~Fig.~1(d) in the companion paper in this
volume~\cite{NSother}) in systems with infinitely many pure states, and a
more complicated structure~\cite{HF87a} in systems with only a single pure
state.  In particular, spin overlap functions computed in systems with
quenched disorder can easily display complicated and nonphysical behavior
that simply reflects the `mismatch' between the boundary condition choice
and the local coupling variables.  Consequently, nontrivial spin overlaps
invariably generated by a change in boundary conditions (e.g., by switching
from periodic to antiperiodic in spin glasses) may carry no more
significance than that, say, in the 2D Ising ferromagnet (for $0<T<T_c$)
generated by Dobrushin boundary conditions. These considerations make it
difficult to believe that a spin overlap variable --- even when confined to
a window --- is likely to uncover generally useful information in
short-range spin glasses.

Nevertheless, one can conceive and construct different, and possibly more
useful, global quantities that can provide useful statistical mechanical
information.  Of greater interest, they {\it can\/} be used in principle to
compare and contrast physically meaningful behaviors of short-range and
infinite-range spin glasses.  We propose here that one such set of
quantities are those related to {\it interfaces\/} between spin
configurations drawn at random from finite-volume Gibbs distributions at
fixed $L$ and $T$.

The interface between two such configurations $\sigma^{L}$ and
$\sigma'^{L}$ is simply the set of all couplings that are satisfied in one
of the two configurations but not the other.  An interface separates
regions where the spins in the two configurations agree from those where
they disagree.  We propose that the global variables of interest are those
--- in particular, density and energy --- characterizing the physical
properties of these interfaces.

The use of interfaces as a probe of spin glass structure is not new.  Our
purpose here is to argue, based on the overall approach described in this
note, that their properties provide a significantly more natural and useful
set of global spin glass variables than the spin overlap function.

The interface density in particular has of course been studied in earlier
papers~\cite{MPRRZ00,NS02,CPPS90,NS2D00,MP00,MP01,NS01}, and its potential
significance has been particularly emphasized
in~\cite{C03,CS04a,CS04b,CG04,CG05}.  The quantity studied is usually the
edge overlap $q_e^{(L)}(\sigma,\sigma')$ between $\sigma^{L}$ and
$\sigma'^{L}$:
\begin{equation}
\label{eq:edge}
q_e^{(L)}(\sigma,\sigma')=N_b^{-1}\sum_{{{x,y\in\Lambda_L}\atop|x-y|=1}}
\sigma_x\sigma_y\sigma'_x\sigma'_y
\end{equation}
where $N_b$ denotes the number of bonds inside $\Lambda_L$.  In the SK
model, the edge overlap is defined similarly, except that the sum runs over
every pair of spins.

In the SK model, there is a trivial relationship between the edge and spin
overlaps.  Consider two spin configurations $\sigma$ and $\sigma'$ in an
$N$-spin system; their spin overlap is defined in the usual way as
\begin{equation}
\label{eq:spin}
q_s^{(N)}(\sigma,\sigma')=N^{-1}\sum_{i=1}^N\sigma_i\sigma'_i\, .
\end{equation}
It is easy to see that 
\begin{equation}
\label{eq:spinedge}
q_e^{(SK)}(\sigma,\sigma')=\Bigl(q_s^{(SK)}(\sigma,\sigma')\Bigr)^2+O(1/N)\,.
\end{equation}
In short-range models, including spin glasses, however, there is no simple
relationship between the two.  For example, the uniform spin configuration
and that with a single domain wall running along the equator (generated,
e.g., using Dobrushin boundary conditions on the ferromagnet at zero
temperature in dimensions greater than or equal to two) have an edge
overlap of one and a spin overlap of zero.

We emphasize, however, that the edge overlap is only one interesting
quantity providing information (in this case, density) about the interface.
We will argue that by itself it does not provide sufficient information to
distinguish among various interesting pictures of the spin glass phase.
Other quantities, in particular the energy scaling of the interface, are
also required for a useful description to emerge.

We can now list several reasons why the density, energy, and possibly other
variables associated with interfaces between states constitute a useful set
of global spin glass variables.

\medskip

0) (Preliminary technical point.) {\it The quantities being studied can be
  clearly defined.\/} This is accomplished through the metastate (see
  Sec.~\ref{sec:meta}) and its natural extensions.  As discussed elsewhere,
  a probability measure on low-energy interfaces can be generated and
  studied through the periodic boundary condition uniform perturbation
  metastate~\cite{NS01}, while one on higher-energy interfaces can be
  constructed via a modification of that used in constructing the {\it
  excitation metastate\/} in~\cite{NS2D00}.

1) {\it The quantities proposed are truly global; i.e., there is no need to
  use a `window'.\/} The reason for this is presented in the next point.

2) {\it The edge overlap, computed in the entire volume, can provide
  unambiguous information about pure state multiplicity.\/} This is in
  contrast to the spin overlap function.  A rigorous formulation and proof
  of this statement (in the case of zero temperature) was provided
  in~\cite{NS02}.  (The results can be extended to nonzero temperatures by
  ``pruning'' from the interfaces small thermally induced
  droplets~\cite{NS02,NS01}.)  Informally stated, the theorem presented in
  that paper stated that if edge overlaps were space-filling (that is,
  their density did not tend to zero with $L$), then there must be multiple
  pure state pairs (e.g., in the appropriate periodic b.c.~metastate).
  Otherwise, there is only a single pair.

It needs to be noted here that this result is restricted to boundary
conditions chosen {\it independently\/} of the couplings (which is always
the case in numerical simulations and theoretical computations).  It is
conceivable that appropriately chosen coupling-{\it dependent\/} boundary
conditions can generate `interface states', separated by domain walls of
vanishing density, as occur below the roughening temperature in Ising
ferromagnets; but no procedure for constructing such boundary conditions
has yet been found.

3) {\it Interface structure also provides information on thermodynamically
  relevant non-pure state structure, in particular, the distribution and
  energies of excitations.\/} This is a potentially important use of the
  procedures of Krzakala-Martin~\cite{KM00} and
  Palassini-Young~\cite{PY00}, and is described in more detail
  in~\cite{NS01}.

4) {\it The scaling of the edge overlap with energy allows one to
   distinguish between different scenarios of the low-temperature phase at
   zero temperature.\/} At zero temperature, the spin overlap function in
   any volume is identical in the
   scaling-droplet~\cite{HF87a,Mac84,BM85,BM87,FH86,FH87b,FH88}, chaotic
   pairs~\cite{NS03jpc,NSBerlin,NS96b,NS97,NS98,NS02,NS92}, and RSB
   scenarios~\cite{MPR97,MPRRZ00,P79,P83,MPSTV84a,MPSTV84b,CPPS90} (see the
   companion paper~\cite{NSother} for a detailed description of these three
   pictures), but the interface density and energy, when used together, can
   distinguish among all of these pictures (as well as the `KM/PY' scenario
   of Krzakala-Martin~\cite{KM00} and Palassini-Young~\cite{PY00}).  This
   is summarized in Fig.~1.

\begin{figure}
\label{fig:table}
\centerline{\epsfig{file=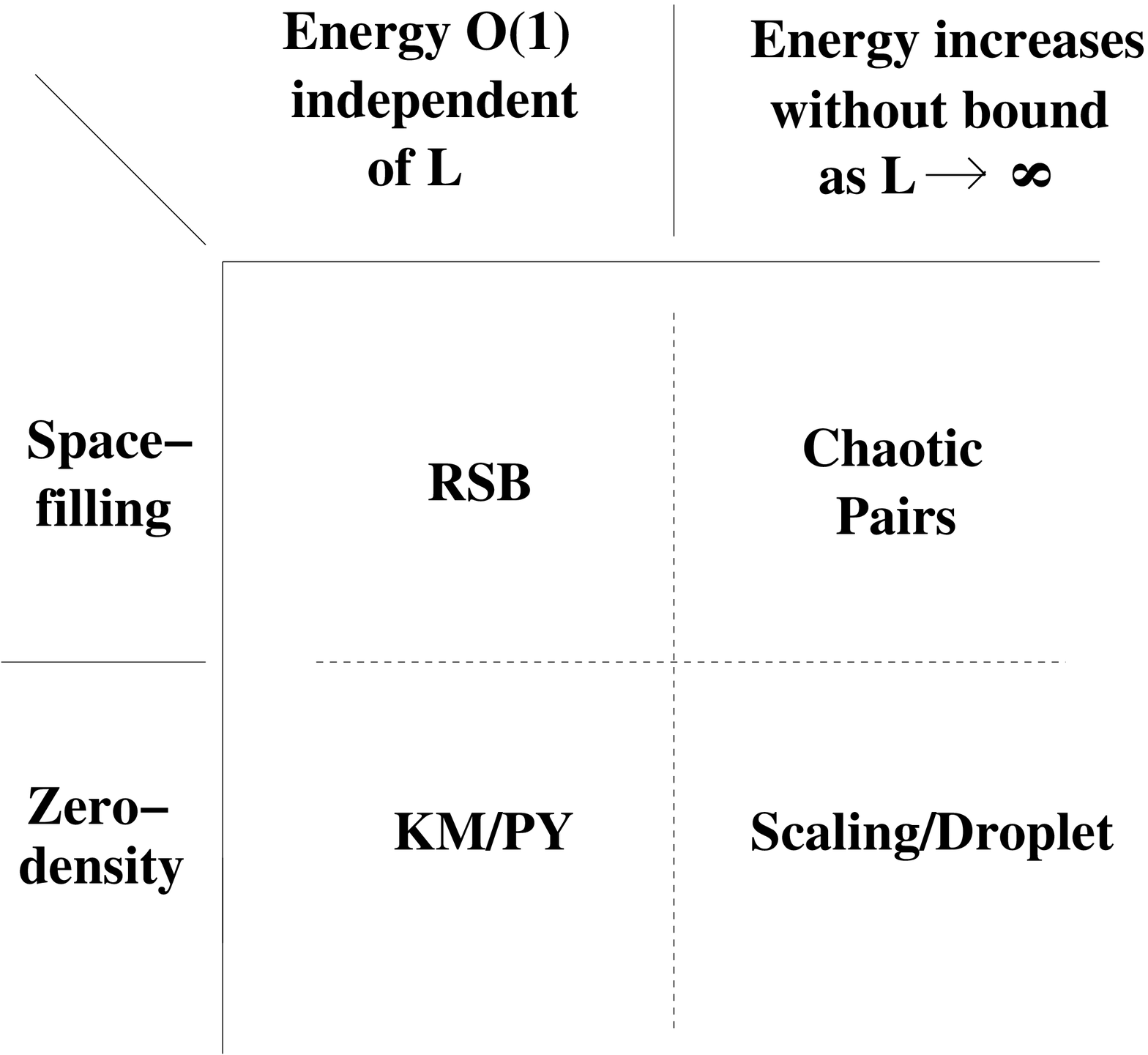,width=3.0in,height=2.5in}}
\caption{
(Adapted from Ref.~\cite{NS02paris}.) 
Table illustrating the correspondence between interface properties and different
scenarios for the structure of the low-temperature phase in short-range spin glasses.}
\end{figure}
\renewcommand{\baselinestretch}{1.25} 
\normalsize

5) {\it The interface properties discussed here provide a means of
  comparing behavior of SK- and short-range spin glasses.\/} That is, use
  of interface properties allows comparison without requiring recourse to
  pure state notions, which as discussed above are poorly defined in the SK
  model.  So, while they provide much (although not all -- see below)
  useful information about pure state structure in short-range models, they
  are also well-defined in SK models and allow for direct comparison of the
  two.

\medskip

It should always be kept in mind, however, that ultimately the interface
variables we have been discussing discard a significant amount of important
information on local correlations; this is to be expected from any global
variable.  These variables should therefore be viewed as providing {\it
additional\/} useful information to the usual local (i.e., thermodynamic)
variables; it is dangerous to view them as a replacement for those
variables.  Consequently, even while providing the arguments in this
section, we emphasize that interface --- or any other global --- variables
can play at best a secondary role in describing the statistical mechanics
of finite-range models, where well-defined state-based quantities already
exist.  It is only in the SK model, where such quantities are mostly
absent, that global quantities play a more primary --- perhaps the only ---
role.

Throughout much of this section we referred to useful properties of a
thermodynamic object we have called the metastate.  We conclude by defining
this object, which in turn enables us to return to the observation made at
the beginning of this note.

\section{What Are Metastates, Anyway?}
\label{sec:meta}

We are interested in the thermodynamic behavior of large, finite systems,
containing on the order of $10^{23}$ interacting degrees of freedom.
Corresponding infinite volume limits serve two purposes.  Mathematically,
they enable precise definitions of quantities corresponding to physical
variables; physically, they allow one to approximate large finite systems
(usually when surface effects can be ignored compared to bulk phenomena).
They allow a deep conceptual understanding of important physical phenomena;
probably the best-known example is understanding phase transitions in terms
of singularities or discontinuities of thermodynamic functions.  It is of
course a fact that such phenomena correspond to true mathematical
singularities only in strictly infinite systems, while common sense
dictates that phase transitions in physical systems are quite real, and
involve behavior as singular or discontinuous as can be found anywhere in
the physical world.  That is precisely why infinite volume limits are
properly regarded as convenient and useful mathematical descriptions of
large finite systems.

When dealing with the usual sort of {\it global\/} variable such as energy,
we see no serious issues arising in disordered systems, nor do we expect
that there is any conceptual divergence from homogeneous systems.  Thus,
for example, the calculation of the SK free energy per spin in the Parisi
solution relies completely on taking the $N\to\infty$ limit (see,
e.g.,~\cite{MPV87}). Any difficulties involved are really of a technical,
rather than a conceptual, nature, and they have finally been rigorously
resolved~\cite{Gu03, Tal}.

It is when dealing with {\it local\/} variables that serious problems
arise, and here the behavior of models with quenched disorder seems to diverge
dramatically from that of homogeneous ones.  In almost all well-known
homogeneous models, such as uniform ferromagnets, there is simple
convergence, as volume goes to infinity, to a single Gibbs state at high
temperature; or to several, related via well-understood symmetry
transformations, at low temperature.  There the nature of the finite-volume
approximation to the infinite-volume limit is conceptually clear and no
difficulties arise.

But what if --- as has often been conjectured for finite-range spin glasses
--- there are many pure states and they are not simply related to each other
by symmetry transformations?  Mathematically, this means that if one
looks at a local variable, such as a single-spin or two-spin correlation
function, there exist many possible (subsequence) limits.  When this
happens, it is not even immediately clear how to obtain
a well-defined  infinite-volume
limit.  This is largely a consequence of {\it chaotic size dependence\/},
first demonstrated in~\cite{NS92} as an unavoidable signature of many states.
Briefly put, local variables, such as correlations, will vary chaotically
and unpredictably as volume size changes (with, say, periodic boundary
conditions on each volume).  In fact, this chaotic size dependence was
proposed by us as a test of the presence of many pure state pairs.
Consequently, convergence of {\it states\/} to a thermodynamic limit is no
longer as simple as just taking a sequence of volumes of arithmetically,
say, or geometrically increasing lengths --- a process that works fine for
homogeneous systems.
 
It turns out that such limits {\it can\/} be defined, but it takes a little
work~\cite{NSBerlin,NS96a}.  However, the existence of limiting
thermodynamic {\it states\/} turns out not to be the essential problem.  As
noted, such states do exist and are well-defined; but they turn out not to
contain the information needed to fully understand the system {\it
in large finite volumes\/}.  So now there does, at first glance, seem to be
a conflict between the thermodynamic limit and behavior in large finite
volumes. But such a conclusion is premature --- with more work, not only
can the two be reconciled again, but an entirely new set of concepts and
insights arises.

So we turn to the question: is it even possible in such systems to describe
the nature of large finite volume systems via a single infinite volume
object, and if so, how?  The answer to the first question is
yes~\cite{NS96b}, and to the second is: by an infinite volume object that
captures the nature of the behavior of the finite systems as volume
increases --- i.e., by the {\it
metastate\/}~\cite{NS03jpc,NSBerlin,NS96b,NS97,NS98,AW90} describing the
empirical distribution of local variables as volume increases without
bound.

Such behavior in $L$ is analogous to chaotic behavior in time $t$ along the
orbit of a chaotic dynamical system.  In each case the behavior is
deterministic but effectively unpredictable.  Consequently, it can be
modelled via random sampling from some distribution $\kappa$ on the space
of states.  In the case of dynamical systems, one can in principle
reconstruct $\kappa$ by keeping a record of the proportion of time the
particle spends in each coarse-grained region of state space.  Similarly,
one can prove \cite{NSBerlin,NS96b} that for inhomogeneous systems like
spin glasses, a similar distribution exists: even in the presence of
chaotic size dependence, the {\it fraction\/} of volumes in which a
given thermodynamic state $\Gamma$ appears, converges (at least along a
sparse sequence of volumes).  By saying that a thermodynamic state $\Gamma$
(which is an infinite-volume quantity) `appears' within a finite volume
$\Lambda_L$, we mean the following: within a window deep inside the volume,
correlation functions computed using the finite-volume Gibbs state $\rho_L$
are the same as those computed using $\Gamma$ (with negligibly small
deviations).

Hence, the metastate allows one to reconstruct the behavior of large finite
volumes from an infinite-volume object, which contains far more information
than any mixed thermodynamic state, such as those often used as a starting
point in RSB analyses.  Because technical details have been provided in
several other places~\cite{NSBerlin,NS96b}, we do not repeat them here.  We
do however refer the reader to~\cite{NSother} where we examine more closely
the nature of the low-temperature spin glass phase, from the point of view
of metastates.

{\it Acknowledgments.}  This research was partially supported by NSF
Grants DMS-01-02587~(CMN) and DMS-01-02541~(DLS).

\renewcommand{\baselinestretch}{1.0} 

\end{document}